# Symbolic Regression Discovery of New Perovskite Catalysts with High Oxygen Evolution Reaction Activity

Baicheng Weng[a,b,c,1], Zhilong Song[b,1], Rilong Zhu[d], Qingyu Yan[d], Qingde Sun[b], Corey G. Grice[a], Yanfa Yan[a,2], Wan-Jian Yin[b,e,2]

[a]Department of Physics & Astronomy, and Wright Center for Photovoltaics Innovation and Commercialization, The University of Toledo, Toledo, OH 43606, USA

[b]Colledge of Energy, Soochow Institute for Energy and Materials InnovationS (SIEMIS), and Jiangsu Provincial Key Laboratory for Advanced Carbon Materials and Wearable Energy Technologies, Soochow University, Suzhou 215006, China

[c]College of Chemistry and Chemical Engineering, Central South University, Changsha 410083, China

[d]College of Chemistry and Chemical Engineering, Hunan University, Changsha 410082, China

[e]Key Lab of Advanced Optical Manufacturing Technologies of Jiangsu Province & Key Lab of Modern Optical Technologies of Education Ministry of China, Soochow University, Suzhou 215006, China

**Symbolic regression (SR) is an emerging method for building analytical formulas to find models that best fit data sets. Here, SR was used to guide the design of new oxide perovskite catalysts with improved oxygen evolution reaction (OER) activities. An unprecedentedly simple descriptor, $\mu/t$, where $\mu$ and $t$ are the octahedral and tolerance factors, respectively, was identified, which accelerated the discovery of a series of new oxide perovskite catalysts with improved OER activity. We successfully synthesized five new oxide perovskites and characterized their OER activities. Remarkably, four of them, $Cs_{0.4}La_{0.6}Mn_{0.25}Co_{0.75}O_3$, $Cs_{0.3}La_{0.7}NiO_3$, $SrNi_{0.75}Co_{0.25}O_3$, and $Sr_{0.25}Ba_{0.75}NiO_3$, outperform the current state-of-the-art oxide perovskite catalyst, $Ba_{0.5}Sr_{0.5}Co_{0.8}Fe_{0.2}O_3$ (BSCF). Our results demonstrate the potential of SR for accelerating data-driven design and discovery of new materials with improved properties.**

Statistical machine learning (ML) is increasingly used in the field of materials informatics as it is an effective tool for discovering quantitative structure– or composition–property relationships and can accelerate materials design (1,2). However, statistical ML does not provide obvious physical insights into the studied data sets, which limits its potential in certain cases. Therefore, statistical ML is not easily accessed by a wide range of researchers, in particular, experimentalists not familiar with theoretical algorithms. Symbolic regression (SR) is an alternative approach that can search for an optimal function with multiple features as variables that describes a given dataset. In contrast to statistical ML, which is a "black-box" method, SR delivers analytical forms that may have physical meanings. Although SR has great potential, its application in the field of material science is still limited (1).

In this paper, we demonstrate that SR is able to construct effective descriptor that can accelerate the materials discovery. Oxide perovskite catalysts have been selected for our study due to two main reasons. First, oxide perovskites ($ABO_3$) are an important family of catalysts for OER applications (6), which are in high demand for renewable energy production and storage, such as hydrogen production from water-splitting (7) and rechargeable metal-air batteries (8), due to their structural flexibility, compositional versatility, and chemical stability (9). Second, the OER activities of oxide perovskite catalysts can be described by descriptors, as demonstrated by various studies over the past sixty years. A number of descriptors, such as the reaction free energy (3,4) and $d$ ($e_g$) orbital filling (5), have been successfully used to understand the trends of OER activities. This has led to the well-known volcano shape of the descriptor-activity curve. However, these descriptors present serious shortfalls. First, the descriptors require dedicated density functional theory (DFT) calculations, which greatly depend on the used methodologies (10). In some cases, it is extremely difficult to calculate these descriptors. For example, accurate determination of $e_g$ occupancy must consider the surface spin state, which is unfortunately not well known (11). Some oxide perovskite surfaces are not well defined due to surface amorphization under OER conditions (12,13). Second, these descriptors do not use any formulas to correlate the OER activity with the descriptor value. The volcano shape of the descriptor–activity curve defines a maximum value from the volcano peak, but it does not quantitatively differentiate the OER activities of oxide perovskite catalysts with the same descriptor values. Therefore, the current descriptors are inherently incapable of predicting and discovering new oxide perovskite catalysts with improved OER activities. An ideal descriptor should be an analytical formula correlating OER activity and the descriptor value. Considering the wide range of A- and B-site dopants and possible compositional combinations, the discovery of new perovskites by only experimental methods is prohibitively slow. Hence, we propose that SR is perfectly suitable for rapid prediction of these materials for specific applications.

Figure 1 shows the workflow diagram of this study. We firstly synthesized a number of well-studied oxide perovskite catalysts to produce consistent and reliable datasets of OER activity for SR analysis. Such data were critical for achieving meaningful formulas (descriptors) that can correlate the materials parameters and OER activities. The OER activity can be quantitatively predicted when the parameters are determined. A simple and accurate descriptor with clear physical insights can help develop strategies to accelerate the discovery of new oxide perovskites. The generality of the descriptor was confirmed by analyzing data reported independently by other research groups. Based on this descriptor, high-throughput screening was conducted to search for new oxide perovskite catalysts with improved OER activities. To validate the predictions, a limited number of new oxide perovskites were synthesized and their OER activities were characterized and compared with their predicted values and current state-of-the-art oxide perovskite catalysts.

The good reliability and comparability of datasets used in SR analysis are of crucial importance for SR in order to produce accurate and insightful formulas (14,15). Since the first discovery of oxide perovskite $LaNiO_3$ as OER catalyst in 1970s (16), chemical management of A- or B-site cations has been used to tune the OER activity due to the structural and chemical flexibility of perovskite structures. The results reported by different groups and produced under different experimental conditions over a period of half a century are summarized in a recent review article (9). To ensure meaningful and valuable SR analysis, we synthesized eighteen known oxide perovskite catalysts (*SI Appendix*, Table. S2; *SI Appendix*, Fig. S4) and characterized their catalytic activity as overpotential vs. RHE (reversible hydrogen electrode), $V_{RHE}$, to build self-consistent, comparable, and reliable datasets. Details of the materials synthesis, along with the structural and electrochemical characterization can be found in the Methods. For $V_{RHE}$ measurements, we selected low disk current densities of 5 $mA{\cdot}cm^{-2}$ in order to avoid other effects, such as conductivity, and normalized the data by the catalyst loading concentration and Brunauer-Emmet-Teller surface area. This ensures that the measured $V_{RHE}$ values are intrinsic values for all oxide perovskite catalysts and hence, comparable. The $V_{RHE}$ values of these eighteen oxide perovskite catalysts measured at 5 $mA{\cdot}cm^{-2}$ disk current are shown in Table 1, where the deviations of the $V_{RHE}$ values are the experimental uncertainty from measurements of at least four different samples with each sample tested at least three times for each composition. Seven of these oxide perovskite catalysts have also been reported by Suntvich *et al.* (6) and the results from both groups showed the same trend in $V_{RHE}$ values (*SI Appendix,* Fig. S6); however, the absolute values are slightly different.

SR was adopted to construct quantitative parameter (descriptor)–activity relationships between the elemental compositions and $V_{RHE}$ of the eighteen oxide perovskites listed in Table 1. The SR process is described in detail in Fig. 2*a* and (*SI Appendix,* Fig. S7-9; *SI Appendix*, Table. S3). To ensure that the SR analysis determines formulas that are insightful for our purpose, it is critical to select relevant parameters to be included in the formulas based on prior knowledge (1). Considering the importance of previous descriptors (3,4,6,17), we chose the following key parameters: the number of *d* electrons for TM ions ($N_d$), tolerance factor $t$ ($\frac{r_A+r_O}{\sqrt{2}(r_B+r_O)}$), octahedral factor $\mu$ ($r_B/r_O$), ionic radii $r_A$ and $r_B$, electronegativity values $\chi_A$ and $\chi_B$, and valence states $Q_A$ and $Q_B$, where A and B refer to the A- and B-site cations, respectively. In the case of the $ABO_3$ perovskite structure, $e_g$ filling is related to $N_d$, $\chi_A$, and $\chi_B$ (6). The stability of perovskite $ABO_3$ is related to $t$ and $\mu$ (18,19) SR initially builds a population of random formulas with the aforementioned parameters as variables. Then, these formulas breed, mutate and evolve to form new ones via genetic programming. The derived formulas compete to model experimental data in the most parsimonious way. Using SR with a grid search of hyper-parameters resulted in about 43,200,000 analytical formulas (descriptor), which were characterized by mean absolute errors (MAE) and complexities, partly described in Fig. 2*b*; see Supplementary Information for more details.

Of the produced descriptors, only those with low MAE (high accuracy) and low complexity are able to provide simple and clear physical insights and be suitable for guiding the high-throughput discovery of new oxide perovskite catalysts. The nine formulas at the Pareto front [marked as A-I in Fig. 2*b*] that met the criteria of simplicity and accuracy among the 43,200,000 candidates are shown in *SI Appendix,* Table. S1. In contrast to conventional descriptors, SR-derived ones are able to quantitatively predict the OER activity and do not require complicated DFT calculations; therefore, the limitations of conventional descriptors are overcome. The predicted $V_{RHE}$ values (marked as diamonds) of the 18 synthesized oxide perovskites using the selected descriptors were in good agreement with the obtained experimental values, as shown in Fig. 2*c* and *SI Appendix,* Fig. S2. All descriptors gave linear and monotonic relationships between $V_{RHE}$ and composition; this is a significant advantage over the conventional volcano-shaped curve (insert of Fig. 2*c*) in terms of predicting new oxide perovskite catalysts with improved OER activity. In principle, all these descriptors could be used to calculate $V_{RHE}$ and search for new oxide perovskite catalysts by screening all compositional combinations of oxide perovskites. In this work, we used a two-parameter formula, i.e., $V_{RHE} = 1.612\mu/t + 1.073$ (eV) (E point at Fig. 2c), to develop strategies to accelerate the discovery of these materials. In this formula, $\mu/t$ has a linear correlation with $V_{RHE}$ and provides the opportunity to tune both the A and B site during materials design. To verify the suitability of this descriptor, we also used $\mu/t$ to fit the experimental data reported in Ref. 6. As shown in the inset of Fig. 2, $\mu/t$ provided a clear linear and monotonic correlation with $V_{RHE}$, confirming the generality of this descriptor. The fitting accuracy of $\mu/t$ was comparable to the volcano shape for descriptor $e_g$ (*SI Appendix,* Fig. S3), although the former provides clear physical insights and guidance for materials design.

The formula $V_{RHE} = 1.612\mu/t + 1.073$ (eV) reveals that the OER activity of oxide perovskite catalysts is closely related to the structural factors of the catalysts, i.e., a smaller $\mu$ and a larger $t$ should lead to higher OER activity. Accordingly, we used a rational strategy to accelerate the screening process: adopting large cations on the A site (increasing $t$) and small cations on the B site (decreasing $\mu$). Previously, the commonly used A-site cations in oxide perovskite catalysts are group IIA (Ca, Sr, Ba) and group IIIB (La, Ce, Pr) elements (9). Based on the insight of the new descriptor developed here, we considered incorporating large group-IA elements (K, Rb, Cs) onto the A site to increase $t$. Among the TM ions that can form perovskite oxides, 3*d* TM ions have the smallest ionic radii, which is consistent with the fact that all existing active oxide perovskite catalysts contain Mn, Fe, Co, and Ni cations (smallest ones among the 3*d* TM ions) on the B site, 4*d*/5*d* TM oxide perovskites are catalytically less active despite having similar *d* electron configurations. Therefore, we considered that the A site contains up to two ions from ($K^{1+}$, $Rb^{1+}$, $Cs^{1+}$, $Ca^{2+}$, $Sr^{2+}$, $Ba^{2+}$, $La^{3+}$, $Ce^{3+}$, $Pr^{3+}$) and the B site contains up to eight ions from ($Mn^{3+}$, $Mn^{4+}$, $Fe^{3+}$, $Fe^{4+}$, $Co^{3+}$, $Co^{4+}$, $Ni^{3+}$, $Ni^{4+}$) with variation in an increment of 0.25 for A and B ionic ratio. Subject to the requirement of charge balance, 3,545 oxide perovskites were obtained from the SR analysis and their $\mu/t$ values were calculated. These oxide perovskites are listed in Data S1 in order or increasing $\mu/t$ value. There are many new oxide perovskites with $\mu/t$ values smaller than those of materials reported in the literature, opening up a new large group of previously unexplored OER catalysts.

We selected thirteen new oxide perovskites in the smallest $\mu/t$ values (the topmost region in Data S1) with an increment of ~0.015 in $\mu/t$ values to consider sufficient elemental and compositional diversity for experimental verification. These thirteen perovskite oxides are: $Ba_{0.75}$ $Sr_{0.25}NiO_3$, $Cs_{0.4}La_{0.6}Mn_{0.25}Co_{0.75}O_3$, $SrNi_{0.75}Co_{0.25}O_3$, $Cs_{0.3}La_{0.7}NiO_3$, $Cs_{0.25}La_{0.75}Mn_{0.5}Ni_{0.5}O_3$, $Cs_{0.5}La_{0.5}Mn_{0.5}Ni_{0.5}O_3$, $Sr_{0.25}La_{0.75}Mn_{0.5}Fe_{0.5}O_3$, $Ba_{0.75}Pr_{0.25}Ni_{0.5}Fe_{0.5}O_3$, $Cs_{0.6}La_{0.4}Mn_{0.75}Co_{0.25}O_3$, $Cs_{0.5}La_{0.5}MnO_3$, $Cs_{0.5}La_{0.5}Mn_{0.25}Co_{0.75}O_3$, $Cs_{0.5}La_{0.5}Mn_{0.5}Co_{0.5}O_3$, and $Cs_{0.25}Pr_{0.75}Mn_{0.25}Fe_{0.25}Co_{0.25}Ni_{0.25}O_3$. The synthesis method is described in detail in the Methods section. We found that eight of them contained significant amounts of impurity or secondary phases, as indicated by the asterisks in the powder X-ray diffraction (PXRD) patterns (*SI Appendix,* Fig. S4). For example, $Cs_{0.5}La_{0.5}Mn_{0.5}Ni_{0.5}O_3$, $Cs_{0.6}La_{0.4}Mn_{0.75}Co_{0.25}O_3$, $Cs_{0.5}La_{0.5}MnO_3$, $Cs_{0.5}La_{0.5}Mn_{0.25}Co_{0.75}O_3$, and $Cs_{0.5}La_{0.5}Mn_{0.5}Co_{0.5}O_3$ showed an impurity phase of $MnO_{4+\delta}$ (main diffraction peaks at 12° and 24°). $Ba_{0.75}Pr_{0.25}Ni_{0.5}Fe_{0.5}O_3$ contained $Pr_2O_3$ and NiO impurity phases. Five compounds including $Cs_{0.4}La_{0.6}Mn_{0.25}Co_{0.75}O_3$, $Cs_{0.3}La_{0.7}NiO_3$, $Cs_{0.25}La_{0.75}Mn_{0.5}Ni_{0.5}O_3$, $Sr_{0.25}Ba_{0.75}NiO_3$, and $SrNi_{0.75}Co_{0.25}O_3$, formed pure perovskite phase, as by confirmed PXRD (*SI Appendix,* Fig. S4). The OER activities of these five new pure oxide perovskites were then characterized (Fig. 3*a-c*). $Cs_{0.4}La_{0.6}Mn_{0.25}Co_{0.75}O_3$, $Cs_{0.3}La_{0.7}NiO_3$, $SrNi_{0.75}Co_{0.25}O_3$, and $Sr_{0.25}Ba_{0.75}NiO_3$ showed lower $V_{RHE}$ values at 5 $mA\cdot cm^{-2}$ (higher OER activity) than BSCF, which is the best oxide perovskite OER catalyst reported in the literature to date. Remarkably, the experimental $V_{RHE}$ values of these new oxide perovskite catalysts were in reasonably good agreement with the values predicted by the SR-derived descriptor, $\mu/t$, as shown in Table 1 and Fig. 2*c*. It is worth noting that we have only selected a very limited number of compositions for experimental synthesis and characterization due to limited resources. It is highly anticipated that more of these predicted oxide perovskite catalysts with high OER activities can be experimentally synthesized and their OER activities will be verified.

The stability of the four new oxide perovskite catalysts with OER activities higher than previously reported oxide perovskite catalysts were tested galvanostatically at 10 $mA\cdot cm^{-2}$ disk current (Fig. 3*d*). We selected a higher disk current density for stability testing to verify the activity decay under strong polarization conditions. $Cs_{0.4}La_{0.6}Mn_{0.25}Co_{0.75}O_3$, $Cs_{0.3}La_{0.7}NiO_3$, $SrNi_{0.75}Co_{0.25}O_3$, and $Sr_{0.25}Ba_{0.75}NiO_3$ showed lower activity degradation than BSCF. In particular, the $Sr_{0.25}Ba_{0.75}NiO_3$ electrode maintained a stable $V_{RHE}$ over 12 h of stability testing without significant decay. Under the same conditions, the BSCF sample showed a much faster degradation rate, with only 90% retention after 9 h. After OER durability tests, the $Sr_{0.25}Ba_{0.75}NiO_3$ electrode maintained its original morphology. Scanning transmission electron microscopy (STEM) and high-resolution transmission electron microscopy images revealed no significant surface amorphization. The surfaces of the $Sr_{0.25}Ba_{0.75}NiO_3$ particles maintained good crystallinity after stability tests, as confirmed by clear observation of the same lattice spacings, along with FFT images (*SI Appendix*, Fig. S5).

Recent work has shown that increasing the valence states of 3*d*-TMs such as Ni and Co from 2+/3+ to 3+/4+ can boost the OER activities of $LaCoO_3$ and $LaNiO_3$ (20). Interestingly, apart from increasing $t$, $Cs^{1+}$ substitution on the A site is a viable route to enhance the valence states of TM B-site ions in oxide perovskites. This correlates with the SR-derived descriptor, $\mu/t$, since increasing the valence states will inevitably reduce the ionic radii of TMs, which will in turn reduce the $\mu$ value, and, therefore, reduce $\mu/t$. Meanwhile, recent theoretical reports predicted that $SrNiO_3$ should have high OER activity (21). Unfortunately, the hexagonal close packing of Sr and O atoms prevents the formation of the perovskite structure. To mitigate this issue, La was proposed to partially substitute Sr. However, partial La substitution leads to the formation of a Ruddlesden–Popper crystal structure instead of perovskite structures (22). Interestingly, the descriptor $\mu/t$ suggests that partial substitution of Sr using larger Ba atoms can enhance catalytic activity. Our experiments showed that $Ba_{0.75}Sr_{0.25}NiO_3$ can be synthesized in the perovskite structure and its OER activity is even higher than BSCF, demonstrating the power of the SR-derived descriptor.

In summary, we used SR to identify a simple descriptor with clear physical insights for describing the OER activity of oxide perovskite catalysts. This simple descriptor can quantitatively predict the OER activity of oxide perovskites and enabled us to rapidly discover a series of new oxide perovskite catalysts with improved OER activities. For proof of concept, we successfully synthesized five oxide perovskites and four of them exhibited OER activities surpassing existing oxide perovskite catalysts reported in the literature. The predicted $V_{RHE}$ values are in good agreement with the experimental values. We anticipate that more of the predicted new oxide perovskite catalysts can be synthesized and their OER activities verified. Our results demonstrate that SR is a powerful ML technique to discover physically meaningful descriptors when sufficient comparable data is available. Such descriptors can accelerate the discovery of new functional materials with enhanced functionalities.

## Materials and Methods

**Symbolic regression.** Symbolic regression analysis using a genetic algorithm was performed using gplearn (23), a python library that extends scikit-learn, a statistical machine learning tool, to symbolic regression. The hyper-parameters used for symbolic regression are listed in *SI Appendix,* Table. S3. The grid search method was used for population size, tournament factor, and parsimony coefficient. The details of the symbolic regression approach are provided as Supplementary Information.

**Primitive function set.** To perform symbolic regression, a functional set is first prepared. To ensure that the final derived descriptors are as simple as possible, we use only simple mathematical operators and functions in the primitive function set, including ($+, -, \times, \div, \sqrt{}$). These primitive functions and terminal sets are the basic building blocks of analytical formulas.

**Terminal set.** The terminal sets include key materials parameters relevant to the catalytic activity (which should be easily accessible), and correspond to the features in statistical machine learning. Previous studies have shown that using terminal sets based on existing human knowledge can help optimize results (1). Therefore, we chose terminal set ($N_d$, $\varepsilon_d$, $\chi_A$, $\chi_B$, $r_A$, $r_B$, $Q_A$, $t$, $\mu$) ($Q_B$ is trivially dependent on $Q_A$ subject to charge balance $Q_A + Q_B = 6$). For double perovskite and perovskite alloys, the arithmetic averages of $N_d$, $\varepsilon_d$, $\chi_A$, $\chi_B$, $Q_A$, $r_A$, and $r_B$ were taken for A- and B-site cations respectively, and $t$ and $\mu$ are calculated based on averaged $r_A$ and $r_B$ values. The values of terminal sets for all perovskites of interest in this work are provided in Table 1, along with the objective values and experimental $V_{RHE}$ values.

**Evolutionary algorithm.** There are three main kinds of evolutional algorithms, including genetic programming (GP), grammar evolution (GE), and analytical programming (AP) for symbolic regression. Here, we chose GP as it is the most popular algorithm and has proved highly successful in materials science for searching stable crystal and interface structures (24,25). The evolution by GP is realized by applying operators such as mutation, recombination, and selection. Evolutionary algorithms are based on the Darwinian theory of biological evolution, in which individuals that fit within a specified system can survive, while those that do not are discarded. The selection rule for individuals is a fitness metric, defined here as the MAE.

**Experimental synthesis of oxide perovskites.** The oxide perovskites were synthesized using a modified Pechini method following by thermal calcination at 850 °C to 1000 °C under dry air/oxygen atmospheres. Briefly, the acetate or nitrate precursors of the perovskite oxides (4 mmol) were mixed in methanol/$H_2O$ (10 mL, 2:1 *v*:*v*), and citric acid (10 mmol) was added to obtain a clear sol. The mixture was dried at 120 °C and the remaining solid was calcinated at 500 °C for 1 h in air. Then, the obtained powder was ground into fine powder and pressed into pellets with a diameter of 15 mm using a hydraulic press at 20 MPa. Finally, the pellets were calcinated at 850 °C to 1000 °C for 6 h under dry air/oxygen atmospheres.

**Crystal structure characterization.** The structure and phase of the synthesized materials were examined by X-ray diffraction (XRD) (Ultima III, Rigaku, Japan) and Raman spectroscopy (Bruker FT Raman Spectrometer with a laser wavelength of 532 nm). The morphology of the films was characterised using transmission electron microscopy (TEM; JEOL 3011, Japan), scanning transmission electron microscopy (STEM; Hitachi HD-2300A, Japan), and high-resolution TEM (HRTEM; Hitachi HD-3010A, Japan). Elemental compositions were determined using energy-dispersive X-ray spectroscopy (EDS; Oxford Instruments, UK) and inductively coupled plasma mass spectrometry (ICP-MS; Thermo Scientific XSeries 2 ICPMS, USA). The catalyst surface area was determined using Brunauer-Emmet-Teller (BET) analysis, using a BELSORP- mini II (BEL. Japan Inc.) under a flow of $N_2$ gas.

**OER characterisation.** OER characterisation was carried out on a glassy carbon rotating disk electrode. First, 2 mg of catalyst was dissolved in 2 mL ethanol and 100 μL Nafion solution was added. Then, the mixture was sonicated for 30 min to form a homogenous mixture. Subsequently, 90 μL of the slurry was loaded onto the surface of a glassy carbon electrode (GCE; 0.196 $cm^2$) and the electrode was dried at room temperature. The electrolyte was purified to remove trace Fe using $Ni(OH)_2$ powder. The electrochemical measurements were performed using a Voltalab PGZ-301 potentiostat/galvanostat (Radiometer Analytical, France), with a platinum foil and a Ag/AgCl electrode used as the counter and reference electrodes, respectively. The loading amount of the catalysts was 0.168 $mg\cdot cm^{-2}$. All potentials were plotted versus the reversible hydrogen electrode (RHE) as $E_{(RHE)} = E_{(Ag/AgCl)} + 0.197 + 0.0591 \times pH$. All linear sweep voltammetry measurements were carried out at a scan rate of 5 $mV\cdot s^{-1}$. All electrochemical measurements were iR compensated (98%). Each sample was synthesized at least four times and each batch was tested at least three times. The error bars denote variations observed from samples synthesis and OER measurements. The stability test was performed using the controlled current electrolysis method. PXRD measurements verified that all of the obtained materials had the perovskite structure.

**Electrochemical characterization.** The OER activities of these new oxide perovskites and BSCF were evaluated in 0.1 M KOH electrolyte at a scan rate of 5 $mV\cdot s^{-1}$. The experimental reference electrode was Ag/AgCl and all potentials in this paper were further converted to the reversible hydrogen electrode (RHE). To evaluate the intrinsic activities, the current densities were normalized by the loading amount and the BET surface areas in order to exclude the increase in current as a result of high loading content and higher surface area. Normalization was carried out according to the expression: *i* ($mA{\cdot}cm^{-2}$ oxide current) = *i* ($mA{\cdot}cm^{-2}$ disk current) ÷ (loading amount ($g{\cdot}cm^{-2}$) × BET surface area ($cm^2{\cdot}g^{-1}$)). Here, *i* ($mA{\cdot}cm^{-2}$ oxide current) was denoted as the normalized specific activity, while *i* ($mA{\cdot}g^{-1}$ oxide current) = *i* ($mA{\cdot}cm^{-2}$ disk current) ÷ (loading amount ($g{\cdot}cm^{-2}$)) refers to the mass activity.

Author Contributions: W.Y. and Y.Y. conceived the idea and supervised the project. W.Y. supervised the theoretical study. Y.Y. supervised the experimental study. W.Y., Y.Y., and Q.S. performed data collection and analysis. Z.S. and W.Y. carried out SR calculations and analysis. B.W., C.G., Z.Z., and Q.Y. performed materials synthesis and electrochemical characterization. W.Y., Y.Y., and B.W. wrote the paper.

[1]B.W. and Z.S. contributed equally to this study.
[2]To whom correspondence may be addressed. Email: yanfa.yan@utoledo.edu or wjyin@suda.edu.cn

Acknowledgment
W.Y. acknowledges funding support from the National Key Research and Development Program of China (grant No. 2016YFB0700700); National Natural Science Foundation of China (grant No. 11674237 and 51602211); Natural Science Foundation of Jiangsu Province of China (grant No. BK20160299); and the Priority Academic Program Development of Jiangsu Higher Education Institutions (PAPD). The theoretical work was carried out at the National Supercomputer Center in Tianjin and the calculations were performed on TianHe-1(A). This paper presents results from an NSF project (grant No. CBET−1433401) under the “NSF 14−15: NSF/DOE Partnership on Advanced Frontiers in Renewable Hydrogen Fuel Production via Solar Water Splitting Technologies” project, which was co-sponsored by the National Science Foundation, Division of Chemical, Bioengineering, Environmental, and Transport Systems (CBET), and the U.S. Department of Energy, Office of Energy Efficiency and Renewable Energy, Fuel Cell Technologies Office.

1. Yiqun Wang, N. W., James M. Rondinelli, Symbolic regression in materials science. *arXiv:1901.04136v2* (2009).
2. Schmidt, M. & Lipson, H, Distilling Free-Form Natural Laws from Experimental Data. *Science* **324**, 81-85 (2009).
3. Bockris, J. O. & Otagawa, T, The Electrocatalysis Of Oxygen Evolution On Perovskites. *Journal of the Electrochemical Society* **131**, 290-302 (1984).
4. Man, I. C. *et al.*, Universality in Oxygen Evolution Electrocatalysis on Oxide Surfaces. *Chemcatchem* **3**, 1159 - 1165 (2011).
5. Hwang, J. *et al.*, Perovskites in catalysis and electrocatalysis. *Science* **358**, 751-756 (2017).
6. Suntivich, J., May, K. J., Gasteiger, H. A., Goodenough, J. B. & Shao-Horn, Y, A Perovskite Oxide Optimized for Oxygen Evolution Catalysis from Molecular Orbital Principles. *Science* **334**, 1383-1385 (2011).
7. Vojvodic, A. & Norskov, J. K, Optimizing Perovskites for the Water-Splitting Reaction. *Science* **334**, 1355-1356 (2011).
8. Seh, Z. W. *et al.*, Combining theory and experiment in electrocatalysis: Insights into materials design. *Science* **355**, eaad4998 (2017).
9. Yin, W.-J. *et al.*, Oxide perovskites, double perovskites and derivatives for electrocatalysis, photocatalysis, and photovoltaics. *Energy & Environmental Science* **12**, 442-462 (2019).
10. Jacobs, R., Hwang, J., Shao-Horn, Y. & Morgan, D, Assessing Correlations of Perovskite Catalytic Performance with Electronic Structure Descriptors. *Chemistry of Materials* **31**, 785-797 (2019).
11. Haverkort, M. W. *et al.*, Spin state transition in LaCoO3 studied using soft x-ray absorption spectroscopy and magnetic circular dichroism. *Physical Review Letters* **97**, 176405 (2006).

12. May, K. J. *et al.*, Influence of Oxygen Evolution during Water Oxidation on the Surface of Perovskite Oxide Catalysts. *Journal of Physical Chemistry Letters* **3**, 3264-3270 (2012).
13. Risch, M. *et al.*, Structural Changes of Cobalt-Based Perovskites upon Water Oxidation Investigated by EXAFS. *Journal of Physical Chemistry C* **117**, 8628-8635 (2013).
14. De Luna, P., Wei, J., Bengio, Y., Aspuru-Guzik, A. & Sargent, E, Use machine learning to find energy materials. *Nature* **552**, 23-25 (2017).
15. Butler, K. T., Davies, D. W., Cartwright, H., Isayev, O. & Walsh, A, Machine learning for molecular and materials science. *Nature* **559**, 547-555 (2018).
16. Meadowcroft, D, Low-cost oxygen electrode material. *Nature* **226**, 847-848 (1970).
17. Hong, W. T., Welsch, R. E. & Shao-Horn, Y, Descriptors of Oxygen-Evolution Activity for Oxides: A Statistical Evaluation. *Journal of Physical Chemistry C* **120**, 78-86 (2016).
18. Goldschmidt, V. M, Die gesetze der krystallochemie. *Naturwissenschaften* **14**, 477-485 (1926).
19. Sun, Q. & Yin, W.-J, Thermodynamic Stability Trend of Cubic Perovskites. *Journal of the American Chemical Society* **139**, 14905-14908 (2017).
20. Weng, B. *et al.*, A layered Na1-xNiyFe1-yO2 double oxide oxygen evolution reaction electrocatalyst for highly efficient water-splitting. *Energy & Environmental Science* **10**, 121-128 (2017).
21. Rong, X., Parolin, J. & Kolpak, A. M, A Fundamental Relationship between Reaction Mechanism and Stability in Metal Oxide Catalysts for Oxygen Evolution. *Acs Catalysis* **6**, 1153-1158 (2016).
22. Forslund, R. P. *et al.*, Exceptional electrocatalytic oxygen evolution via tunable charge transfer interactions in $La_{0.5}Sr_{1.5}Ni_{1-x}Fe_xO_{4\pm\delta}$ Ruddlesden-Popper oxides. *Nature Communications* **9**, 3150 (2018).
23. *https://gplearn.readthedocs.io/en/latest/intro.html*.
24. Chua, A. L. S., Benedek, N. A., Chen, L., Finnis, M. W. & Sutton, A. P, A genetic algorithm for predicting the structures of interfaces in multicomponent systems. *Nature Materials* **9**, 418-422 (2010).
25. Wu, S. Q. *et al.*, An adaptive genetic algorithm for crystal structure prediction. *Journal of Physics-Condensed Matter* **26**, 035402 (2014).

**Table 1. Key materials parameters and $V_{RHE}$ values of 23 selected oxide perovskites.** The key materials parameters include the tolerance factor ($t$), octahedral factor ($\mu$), ionic radii of A-site ($R_A$) and B-site ($R_B$), electronegativity of A-site ($\chi_A$) and B-site ($\chi_B$), valence state of A-site ($Q_A$), and number of $d$ electrons on TM B-site ($N_d$). The materials are ordered by the value of $\mu/t$ in each dataset of known and new perovskites.

| No. | Materials | $t$ | $\mu$ | $R_A$(Å) | $R_B$(Å) | $\chi_A$ | $\chi_B$ | $Q_A$ | $N_d$ | $\mu/t$ | $V_{RHE}$ (predict) | $V_{RHE}$ (low) | $V_{RHE}$ (high) | $V_{RHE}$ (average) |
|---|---|---|---|---|---|---|---|---|---|---|---|---|---|---|
| **Known Perovskites** | | | | | | | | | | | | | | |
| **1** | $LaMnO_3$ | 0.993 | 0.430 | 1.36 | 0.58 | 1.1 | 1.55 | 3 | 4 | 0.433 | 1.773 | 1.770 | 1.806 | 1.788 |
| **2** | $LaMn_{0.5}Ni_{0.5}O_3$ | 0.998 | 0.422 | 1.36 | 0.57 | 1.1 | 1.73 | 3 | 5.5 | 0.423 | 1.758 | 1.712 | 1.732 | 1.722 |
| **3** | $LaNiO_3$ | 1.003 | 0.415 | 1.36 | 0.56 | 1.1 | 1.91 | 3 | 7 | 0.413 | 1.743 | 1.690 | 1.724 | 1.707 |
| **4** | $LaMn_{0.5}Cu_{0.5}O_3$ | 0.988 | 0.437 | 1.36 | 0.59 | 1.1 | 1.725 | 3 | 6 | 0.442 | 1.789 | 1.762 | 1.791 | 1.774 |
| **5** | $LaNi_{0.9}Fe_{0.1}O_3$ | 1.004 | 0.414 | 1.36 | 0.559 | 1.1 | 1.902 | 3 | 6.8 | 0.413 | 1.742 | 1.770 | 1.809 | 1.790 |
| **6** | $LaNi_{0.8}Fe_{0.2}O_3$ | 1.004 | 0.413 | 1.36 | 0.558 | 1.1 | 1.894 | 3 | 6.6 | 0.412 | 1.740 | 1.737 | 1.780 | 1.759 |
| **7** | $LaFeO_3$ | 1.009 | 0.407 | 1.36 | 0.55 | 1.1 | 1.83 | 3 | 5 | 0.404 | 1.728 | 1.750 | 1.766 | 1.758 |
| **8** | $La_{0.5}Pr_{0.5}FeO_3$ | 1.010 | 0.407 | 1.365 | 0.55 | 1.115 | 1.83 | 3 | 5 | 0.403 | 1.727 | 1.716 | 1.733 | 1.725 |
| **9** | $PrFeO_3$ | 1.012 | 0.407 | 1.37 | 0.55 | 1.13 | 1.83 | 3 | 5 | 0.402 | 1.726 | 1.752 | 1.763 | 1.758 |
| **10** | $LaCoO_3$ | 1.011 | 0.404 | 1.36 | 0.545 | 1.1 | 1.88 | 3 | 6 | 0.399 | 1.721 | 1.708 | 1.734 | 1.721 |
| **11** | $La_{0.5}Ca_{0.5}CoO_3$ | 1.011 | 0.398 | 1.35 | 0.538 | 1.05 | 1.88 | 2.5 | 5.5 | 0.394 | 1.712 | 1.674 | 1.688 | 1.682 |
| **12** | $La_{0.8}Sr_{0.2}CoO_3$ | 1.019 | 0.401 | 1.376 | 0.542 | 1.07 | 1.88 | 2.8 | 5.8 | 0.394 | 1.713 | 1.677 | 1.699 | 1.688 |
| **13** | $Sr_{0.25}La_{0.75}Fe_{0.5}Co_{0.5}O_3$ | 1.020 | 0.401 | 1.38 | 0.542 | 1.063 | 1.855 | 2.75 | 5.25 | 0.393 | 1.711 | 1.695 | 1.741 | 1.718 |
| **14** | $La_{0.4}Sr_{0.6}CoO_3$ | 1.034 | 0.397 | 1.408 | 0.536 | 1.01 | 1.88 | 2.4 | 5.4 | 0.384 | 1.697 | 1.660 | 1.730 | 1.695 |
| **15** | $La_{0.2}Sr_{0.8}CoO_3$ | 1.042 | 0.395 | 1.424 | 0.533 | 0.98 | 1.88 | 2.2 | 5.2 | 0.379 | 1.689 | 1.661 | 1.700 | 1.681 |
| **16** | $SrCoO_3$ | 1.049 | 0.393 | 1.44 | 0.53 | 0.95 | 1.88 | 2 | 5 | 0.374 | 1.682 | 1.658 | 1.682 | 1.670 |
| **17** | $Ba_{0.5}Sr_{0.5}Co_{0.8}Fe_{0.2}O_3$ | 1.082 | 0.391 | 1.525 | 0.528 | 0.92 | 1.876 | 2 | 4.8 | 0.361 | 1.661 | 1.614 | 1.663 | 1.639 |
| **18** | $BaFeO_3$ | 1.119 | 0.385 | 1.61 | 0.52 | 0.89 | 1.83 | 2 | 4 | 0.344 | 1.635 | 1.676 | 1.696 | 1.686 |
| **New Perovskites** | | | | | | | | | | | | | | |
| **19** | $Cs_{0.25}La_{0.75}Mn_{0.5}Ni_{0.5}O_3$ | 1.064 | 0.398 | 1.49 | 0.538 | 1.023 | 1.73 | 2.5 | 5 | 0.374 | 1.682 | 1.662 | 1.683 | 1.673 |
| **20** | $Cs_{0.4}La_{0.6}Mn_{0.25}Co_{0.75}O_3$ | 1.095 | 0.395 | 1.568 | 0.534 | 0.976 | 1.798 | 2.2 | 4.7 | 0.361 | 1.660 | 1.562 | 1.623 | 1.593 |
| **21** | $Cs_{0.3}La_{0.7}NiO_3$ | 1.088 | 0.379 | 1.516 | 0.512 | 1.007 | 1.91 | 2.4 | 6.4 | 0.348 | 1.641 | 1.609 | 1.633 | 1.621 |
| **22** | $SrNi_{0.75}Co_{0.25}O_3$ | 1.071 | 0.365 | 1.44 | 0.493 | 0.95 | 1.903 | 2 | 5.75 | 0.341 | 1.629 | 1.605 | 1.635 | 1.620 |
| **23** | $Sr_{0.25}Ba_{0.75}NiO_3$ | 1.127 | 0.356 | 1.568 | 0.48 | 0.905 | 1.91 | 2 | 6 | 0.315 | 1.589 | 1.568 | 1.632 | 1.600 |

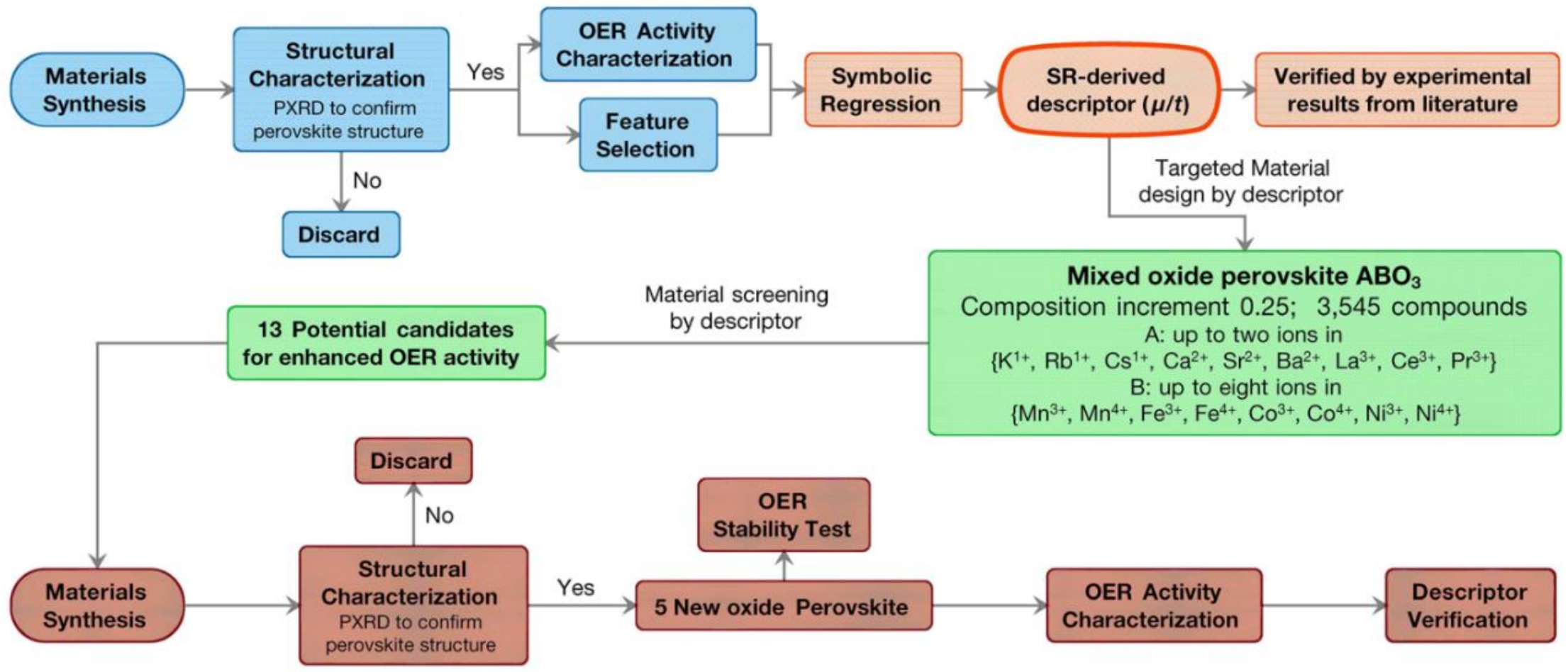

**Fig.1. Workflow diagram.** It contains four major parts: dataset generation (blue), SR (red), materials design and screening (green), experimental verification (brown).

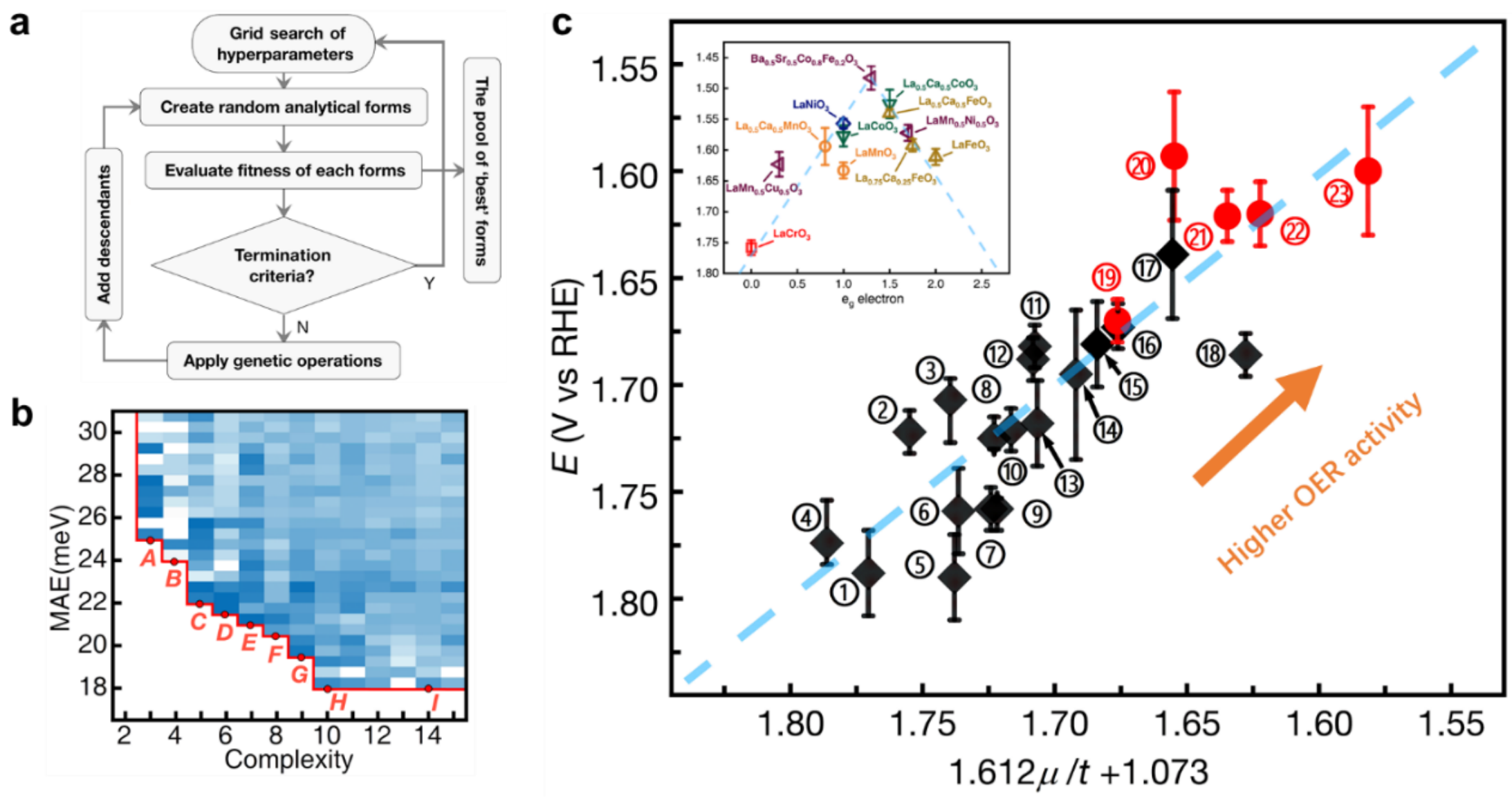

**Fig. 2. The scheme and results of symbolic regression. a,** the flowchart of symbolic regression based on genetic programming (see more details in Supplementary Fig. S3). **b,** Pareto front of MAE vs. complexity of 43,200,000 analytical forms (8,640 generations × 5,000 individuals) shown via density plot. **c,** $V_{RHE}$ vs. 1.612$\mu/t$ + 1.073 (black diamonds: conventional perovskites; red dots: newly-discovered perovskites). The current densities are normalized by BET surface areas and loading amount. For BSCF, the BET surface area is 0.30 $m^2$ $g^{-1}$. Thus, 10 mA $cm^{-2}$ normalized current densities corresponds to 5.0 mA $cm^{-2}$ disc current density for BSCF, and other perovskite oxides are normalized at the same current densities. Therefore, lower potentials mean higher activities. The data details of calculating $\mu/t$ can be found in Data S2. The inset Figure are from Ref. [6] with permissions.

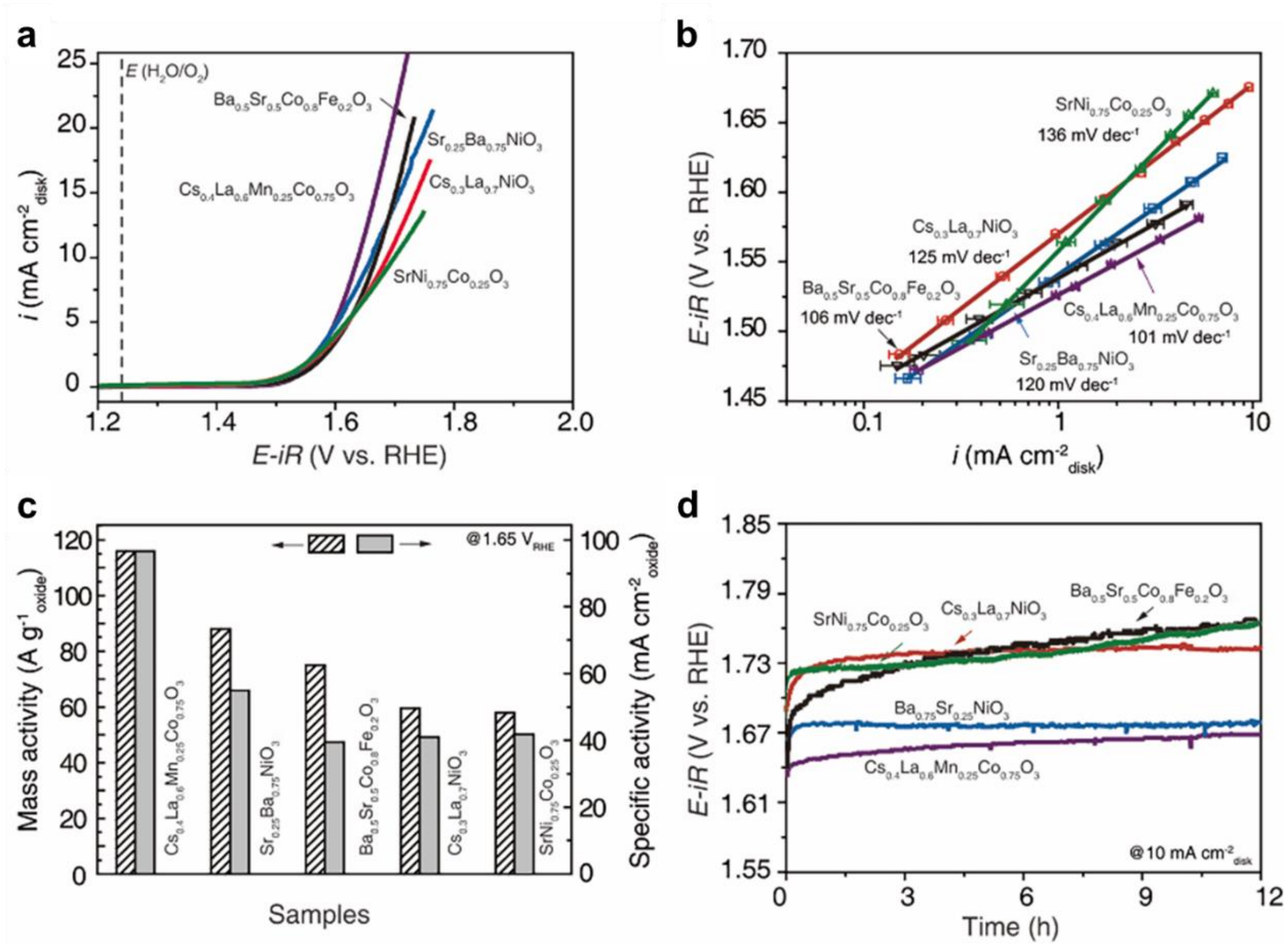

**Fig. 3. Electrochemical characterizations of $Ba_{0.5}Sr_{0.5}Co_{0.8}Fe_{0.2}O_3$ and predicted new oxide perovskites. a,** LSV curves. **b,** corresponding Tafel slopes. **c,** mass and specific activities. **d, results of** stability tests under galvanostatic conditions at 10 $mA{\cdot}cm^{-2}$ disk current density.

# Supplementary Information for

## Symbolic Regression Discovery of New Perovskite Catalysts with High Oxygen Evolution Reaction Activity

Baicheng Weng[a,b,c,1], Zhilong Song[b,1], Rilong Zhu[d], Qingyu Yan[d], Qingde Sun[b], Corey G. Grice[a], Yanfa Yan[a,2], Wan-Jian Yin[b,e,2]

[a]Department of Physics & Astronomy, and Wright Center for Photovoltaics Innovation and Commercialization, The University of Toledo, Toledo, OH 43606, USA

[b]Colledge of Energy, Soochow Institute for Energy and Materials InnovationS (SIEMIS), and Jiangsu Provincial Key Laboratory for Advanced Carbon Materials and Wearable Energy Technologies, Soochow University, Suzhou 215006, China

[c]College of Chemistry and Chemical Engineering, Central South University, Changsha 410083, China

[d]College of Chemistry and Chemical Engineering, Hunan University, Changsha 410082, China

[e]Key Lab of Advanced Optical Manufacturing Technologies of Jiangsu Province & Key Lab of Modern Optical Technologies of Education Ministry of China, Soochow University, Suzhou 215006, China

**A brief introduction to symbolic regression**. Symbolic regression (1-4) is a unique machine learning approach. It is different from statistical machine-learning approach (5) which bears a hidden black-box model and is difficult for physical interpretation. Symbolic regression is to build straightforward and effective descriptors that are able to link the easily-accessed materials parameters with complicated catalytic activities. There are three essential parts for symbolic regression to derive descriptors: primitive function, terminal, and evolutionary algorithm. Their definitions are provided in Method part. In symbolic regression based on genetic programming, an analytical form can be expressed as a tree structure which is composed of primitive functions and terminals (see examples in Fig. S7). The complexity of an analytical form is defined as the number of composed primitive functions and terminals, shown as nodes in Fig. S7.

**Symbolic regression flow chart.** The flow chart of symbolic regression used in current work is shown in Fig. S8. It initially builds a population ($N_{ind}$ =5000) of random analytical formulas composed of primitive functions and terminals, i.e., a random tree structure with random nodes, to represent relationships between materials parameters and catalytic activity ($V_{RHE}$). The performance of formula is measured by mean absolute errors (MAE) between predicted and experimental $V_{RHE}$s for eighteen known perovskites. The algorithm selects the best formula (least MAE) into the pool of final solutions set. To generate the next generation, a part of formulas (here 1000 formulas) are selected by using a tournament method (with tournament size 20) as implemented in gplearn (6). Generic operations of crossover and mutation are then performed among them to form 1000 new ones for the next generations. Examples of crossover and mutation operations are shown in Fig. S7. Another 4000 random formulas are then added to supplement new generation up to totally 5000 formulas. The best formula in new generation is then selected into the final solution set. In principles, the procedure continues until a good formula with desired function metric (MAE < 0.01 eV) is found or the maximum generation reaches $N_{maxG}$= 20. The applicability of genetic algorithm for searching optimal analytical formula is inspired by Darwin's theory of *'natural selection'*.

**Search method for parameters.** For genetic operations, the results may largely depend on how crossover and mutation are performed on 1000 selected formulas in each generation. To mitigate the impact of artificial hyper-parameters on final results, we used grid search of hyper-parameters. The associated parameters for parsimony coefficient, crossover and subtree mutation probability are shown in Table S3. Accordingly, there are 432 sets of hyper-parameters. In each set, there are maximum 20 generations and each generation produces one best individual, which results in about 8,640 individuals. The Pareto front, showing the trade-off between the MAE and complexity, of total 43,200,000 individuals (8640 generations × 5000 individuals) is shown in Fig. 2(c).

**Further analysis of the SR-derived descriptors.** To further analyze the SR-derived descriptors, the analytical forms of nine formulas at Pareto front in Fig. 2(b), are shown in Table S1. It is seen that when only one parameter is chosen, it chooses $t$, $\chi_A$ or $Q_A$. $t$ is a well-known descriptor for perovskite stability. Therefore, the results underlines the importance of stability and A-site cation to the catalytic activity. Among nice formulas, a two-parameter form $\mu/t$ demonstrates a linear correlation with $V_{RHE}$, as clearly shown in Fig. 2(c). This unprecedently simple descriptor is significantly better than the conventional ones since it quantitatively predicts the OER activity for a given $\mu/t$ value. Therefore, in this work, we choose $\mu/t$ as a descriptor for materials design and screening. Notably, the formulas at Pareto front are consistent with each other. They all indicate that A-site cation with large size, low electronegativity and low valence charge should have low $V_{RHE}$. All those indications point to group-IA elements with large size. The accuracy of analytical formula $\mu/t$ can be further improved by increasing formulas' parameters and complexity, as shown in other points at Pareto front. They have marginal effects on fitness metrics but significantly increase the complexity therefore not applicable for high-throughput materials design.

Previous descriptors of reaction free energy and $d$ ($e_g$) orbital filling describe OER activity from the point of view of energetics and electronic properties of $d$ orbitals of TMs. Elemental electronegativity and number of TM $d$ electrons are closely related to $d/e_g$ orbital filling. Though these parameters were included in the SR analysis, simple descriptor $\mu/t$ survived in the *'nature selection'* process in genetic algorithm, suggesting that the previous descriptors may not be the most suitable ones to describe oxide perovskite catalysts.

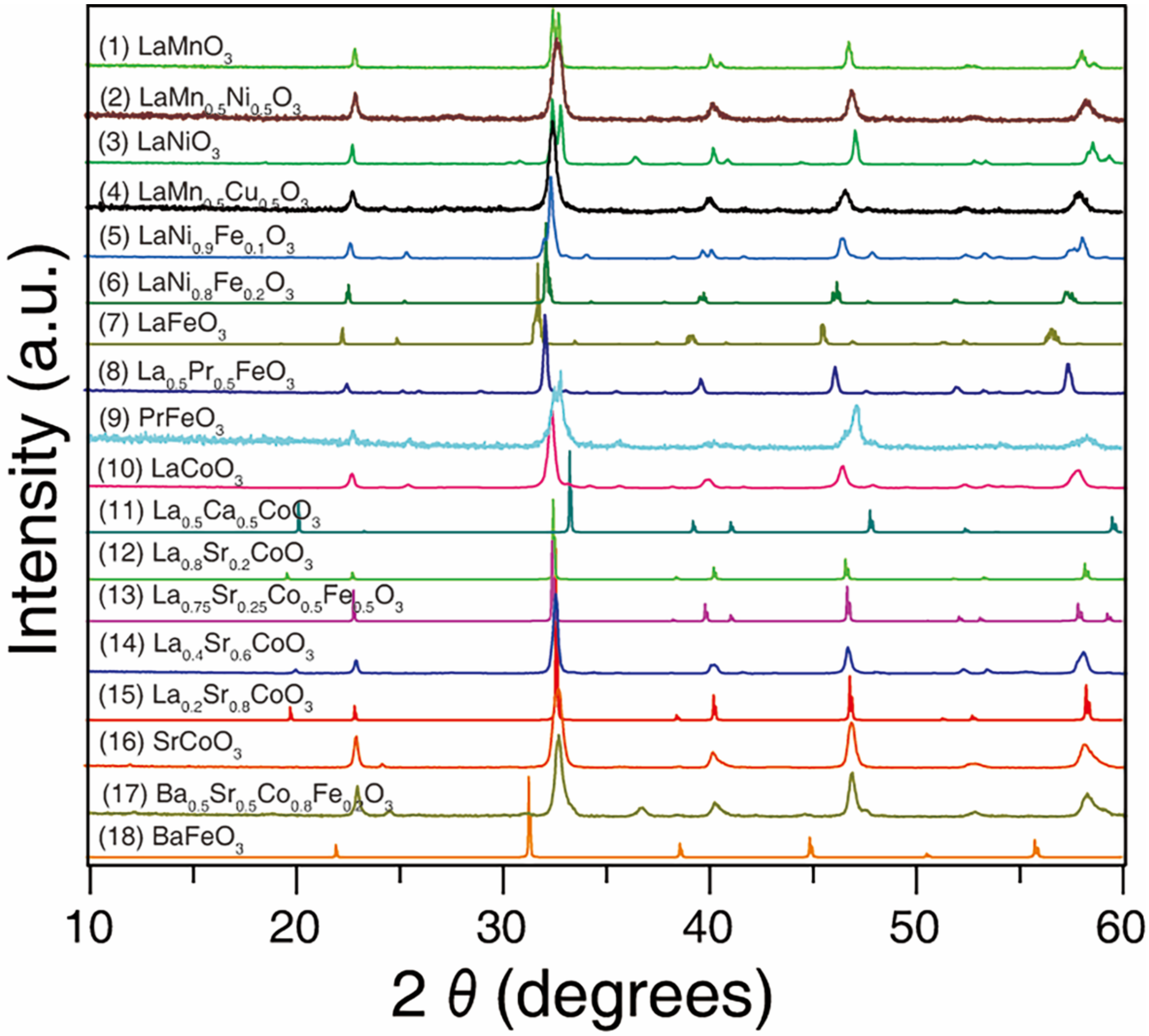

**Fig. S1. PXRD data of 23 oxide perovskite oxides including 18 known oxide perovskites and 5 new ones.** The first characteristic peak of a perovskite structure around 20–25° corresponds to (100) facet, the main peak at 30–35° corresponds to (110) facet and 40° corresponds to (111) facet, 45–50° and 55–60° correspond to (200) and (211) facets, respectively. Here we cannot exclude the possible existence of oxygen vacancies that usually exist in oxide perovskite. For clarity, stoichiometric chemical formulas are adopted in this work.

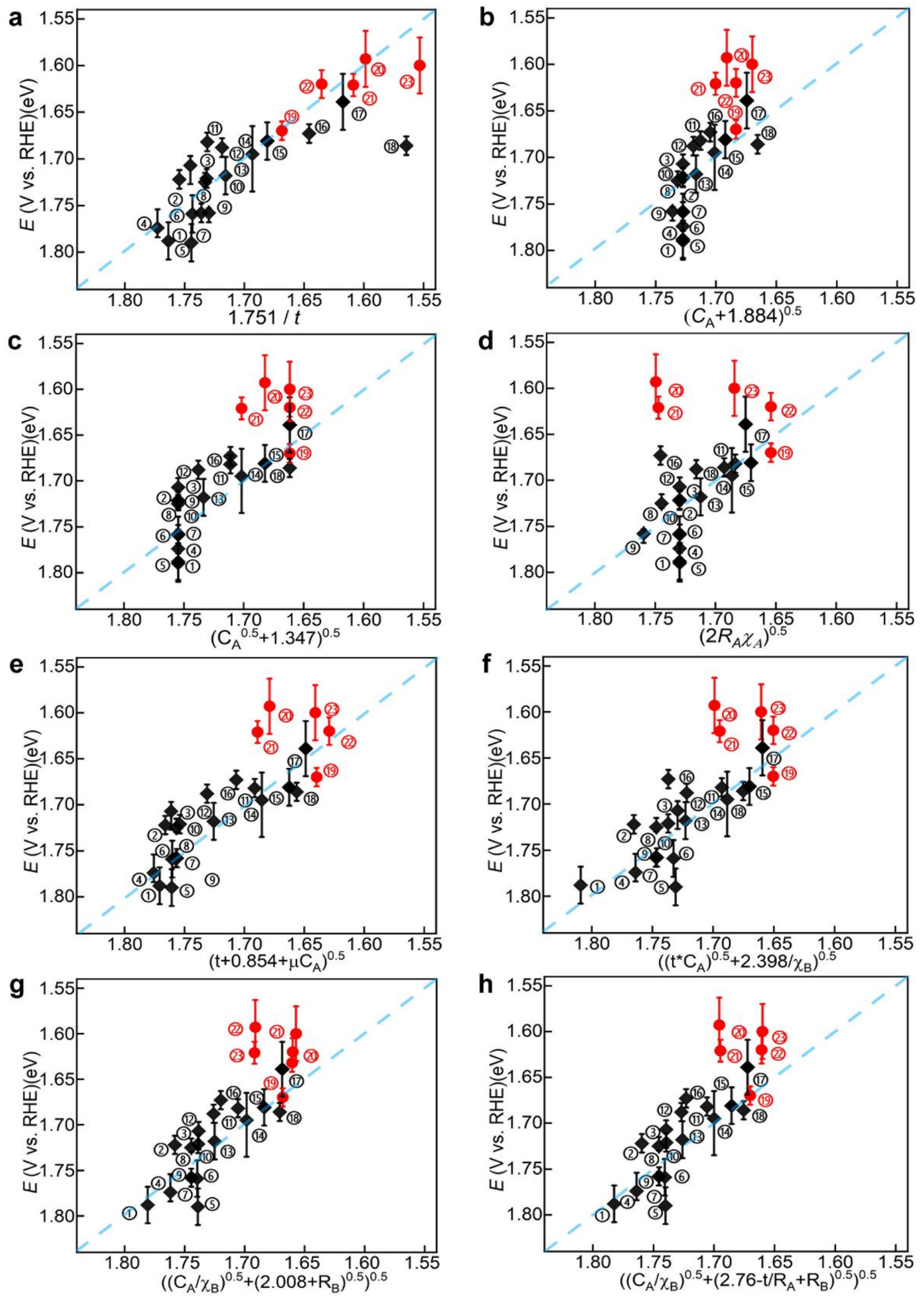

**Fig. S2. The performance of eight descriptors at the Pareto front in Figure 2(c). a,** A formula. **b,** B formula. **c,** C formula. **d,** D formula. **e,** F formula. **f,** G formula. **g,** H formula. **h,** I formula.

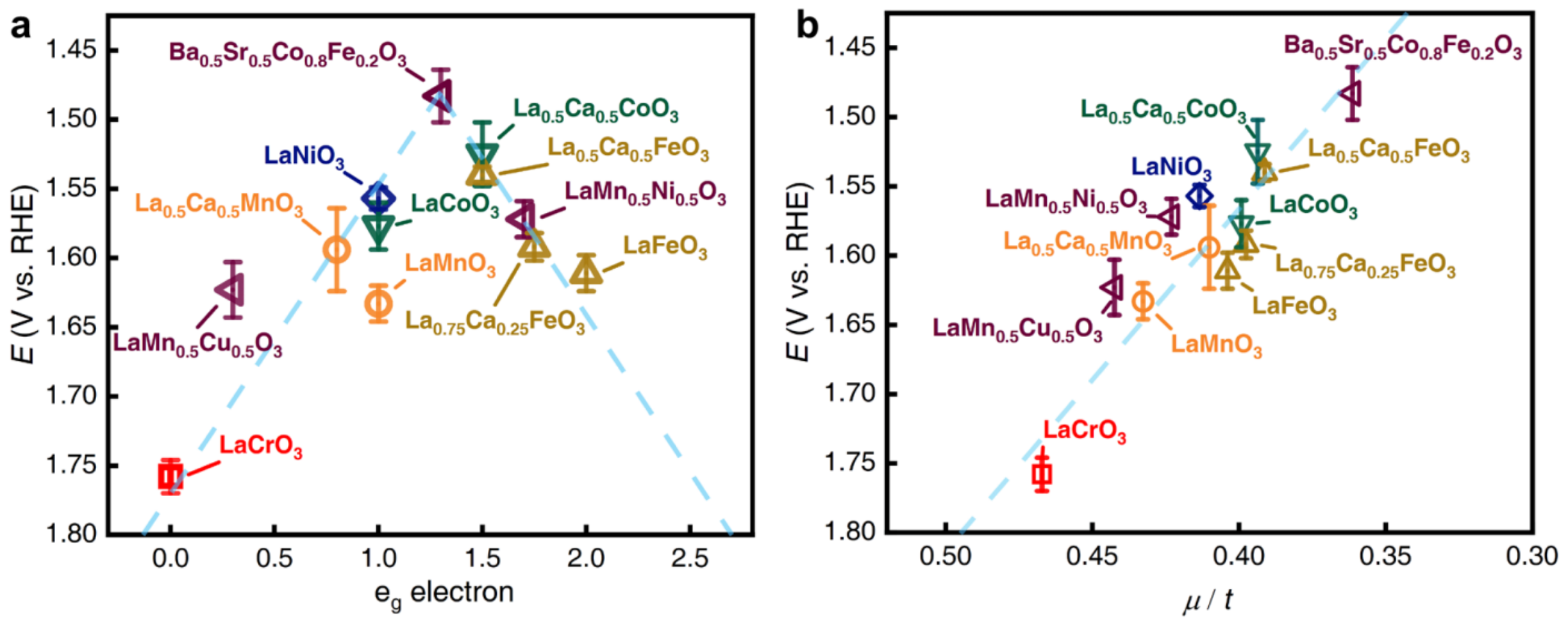

**Fig. S3. Comparison of $e_g$ and $\mu/t$ descriptors based on independent experimental data. a,** Fig. 2 from the study of Suntivich *et al* (6) reproduced with permission from the American Association for the Advancement of Science. **b,** Reformatted plot according to descriptor $\mu/t$. The MAE (Pearson correlation coefficient) for **a** and **b** were 20.6 meV (0.923) and 21.0 meV (0.928) respectively.

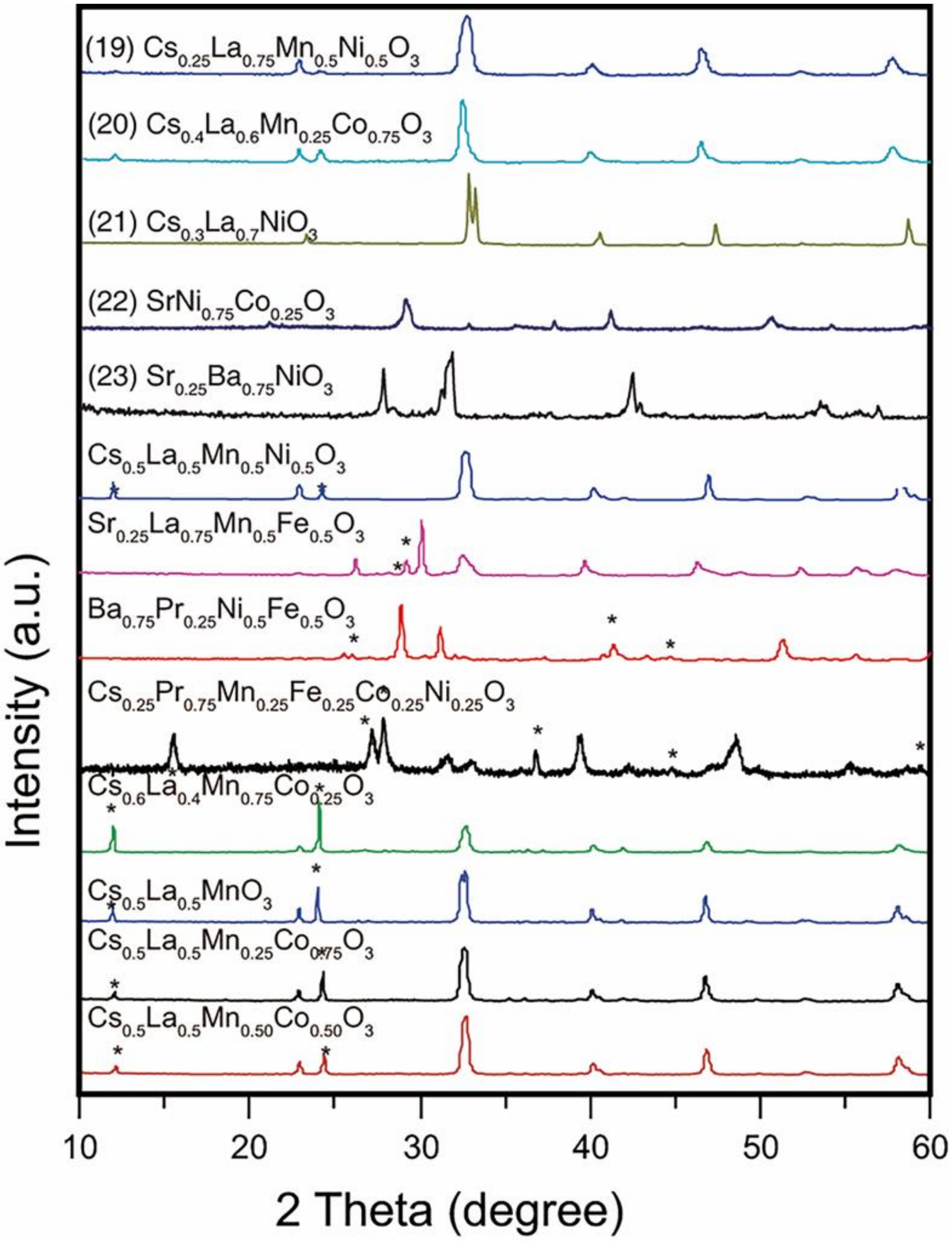

**Fig. S4. PXRD patterns of synthesized oxide perovskite samples containing impurities phases.** The impurities are labelled by asterisks. Although all samples show mainly perovskite structures, there remained a significant amount of impurities that were difficult to remove. For example, the peaks at 12° and 24° correspond to the (040) and (111) facets of $MnO_x$, respectively, and the peak intensities increased with increasing content of Mn and Cs. We speculated that an increase in Cs content may destabilize the structure. The large Mn cations may not easily produce a stable perovskite structure with Cs and La, while the same should be true for Pr with Ba and Cs, due to the large size difference between Pr and Ba/Cs.

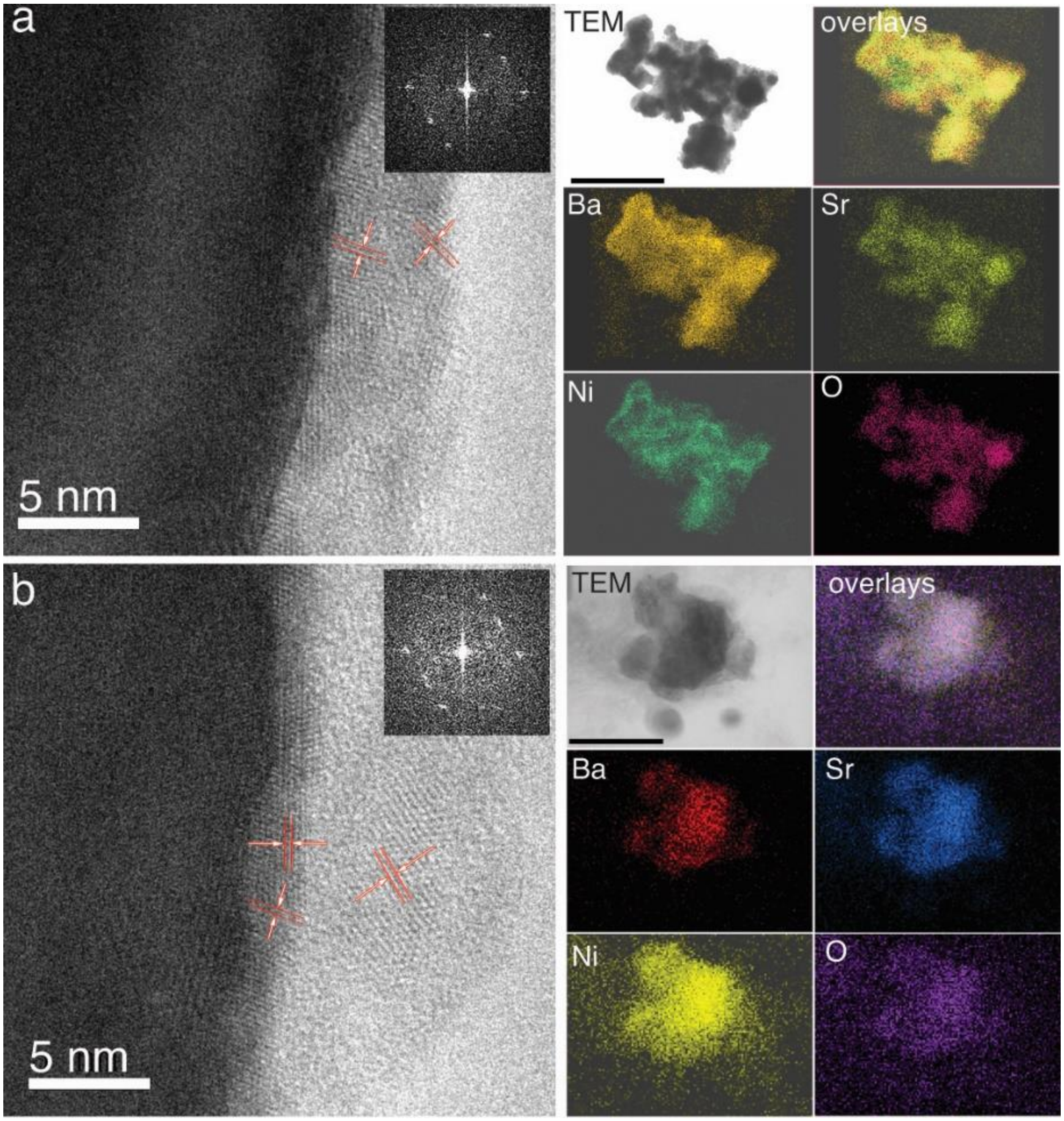

**Fig. S5. Morphology measurements of $Ba_{0.75}Sr_{0.25}NiO_3$ before and after OER testing. a,** HRTEM before a stability test. **b**, HRTEM after a stability test. Right side: STEM atomic mapping (scale bar: 500 nm). The labelled lattice spacing is around 0.3 nm, which corresponded to the (110) lattice planes of $Ba_{0.75}Sr_{0.25}NiO_3$, in good agreement with the PXRD measurements. The insets of (a) and (b) show the fast Fourier transform image of the corresponding HRTEM image. The well-regulated arrayed spots indicated that the grown crystal had high crystallinity. HRTEM of $Ba_{0.75}Sr_{0.25}NiO_3$ before and after OER testing clearly showed the same lattice spacing and very similar fast Fourier transform images, suggesting outstanding stability of the $Ba_{0.75}Sr_{0.25}NiO_3$ sample under OER conditions. It is known that the surface of BSCF after OER testing becomes amorphous, which results in activity decay. The maintenance of good crystallinity indicates that $Ba_{0.75}Sr_{0.25}NiO_3$ is a stable OER electrocatalyst. In order to verify the atomic distribution, STEM mapping was conducted; the even distribution of the atoms over the analyzed area further demonstrates the excellent stability of the sample.

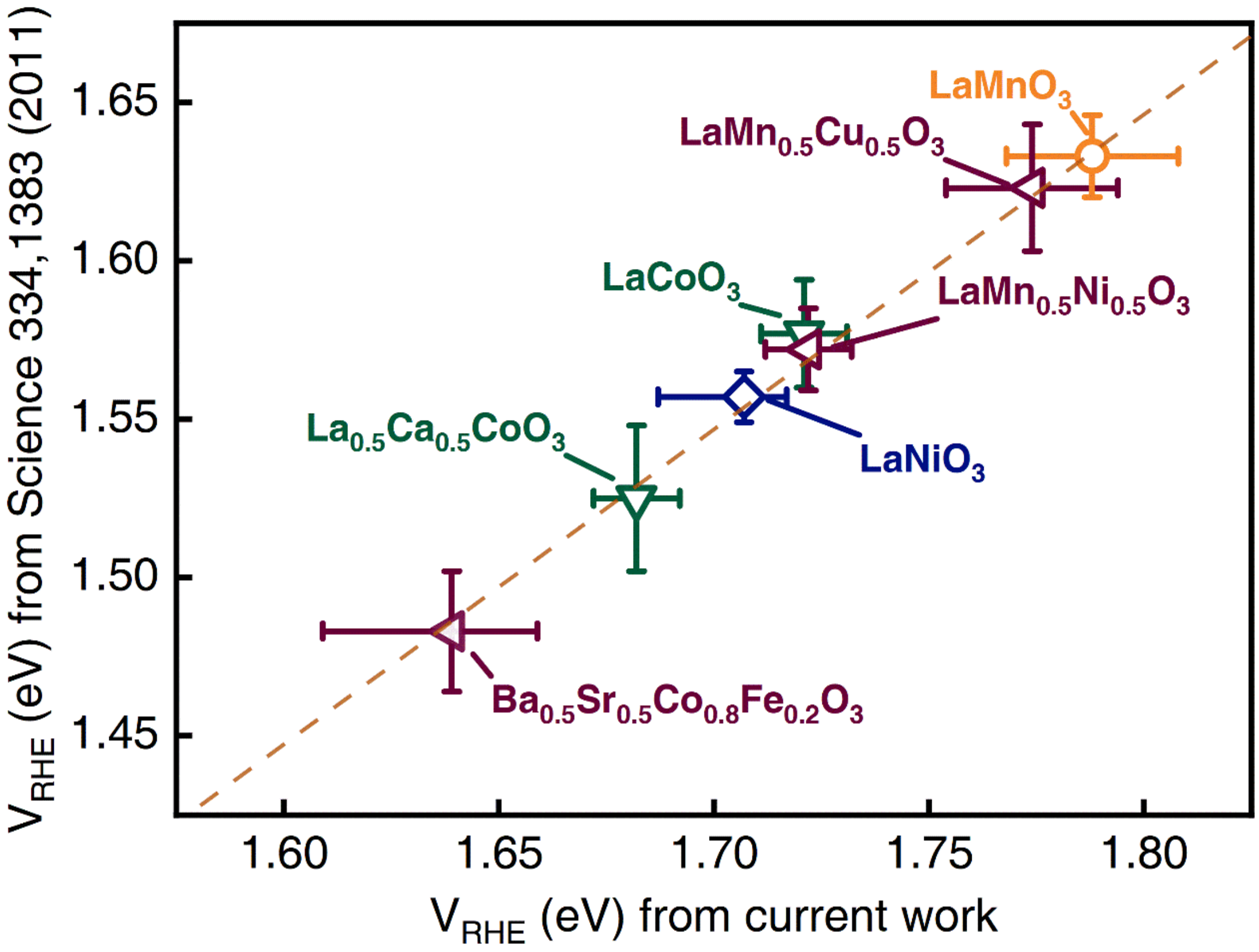

**Fig. S6. Comparison of $V_{RHE}$s of seven oxide perovskite OER catalysts.** The results from this work are presented in the horizon axis and the results reported by Suntivich *et al.* (6) are presented in the vertical axis.

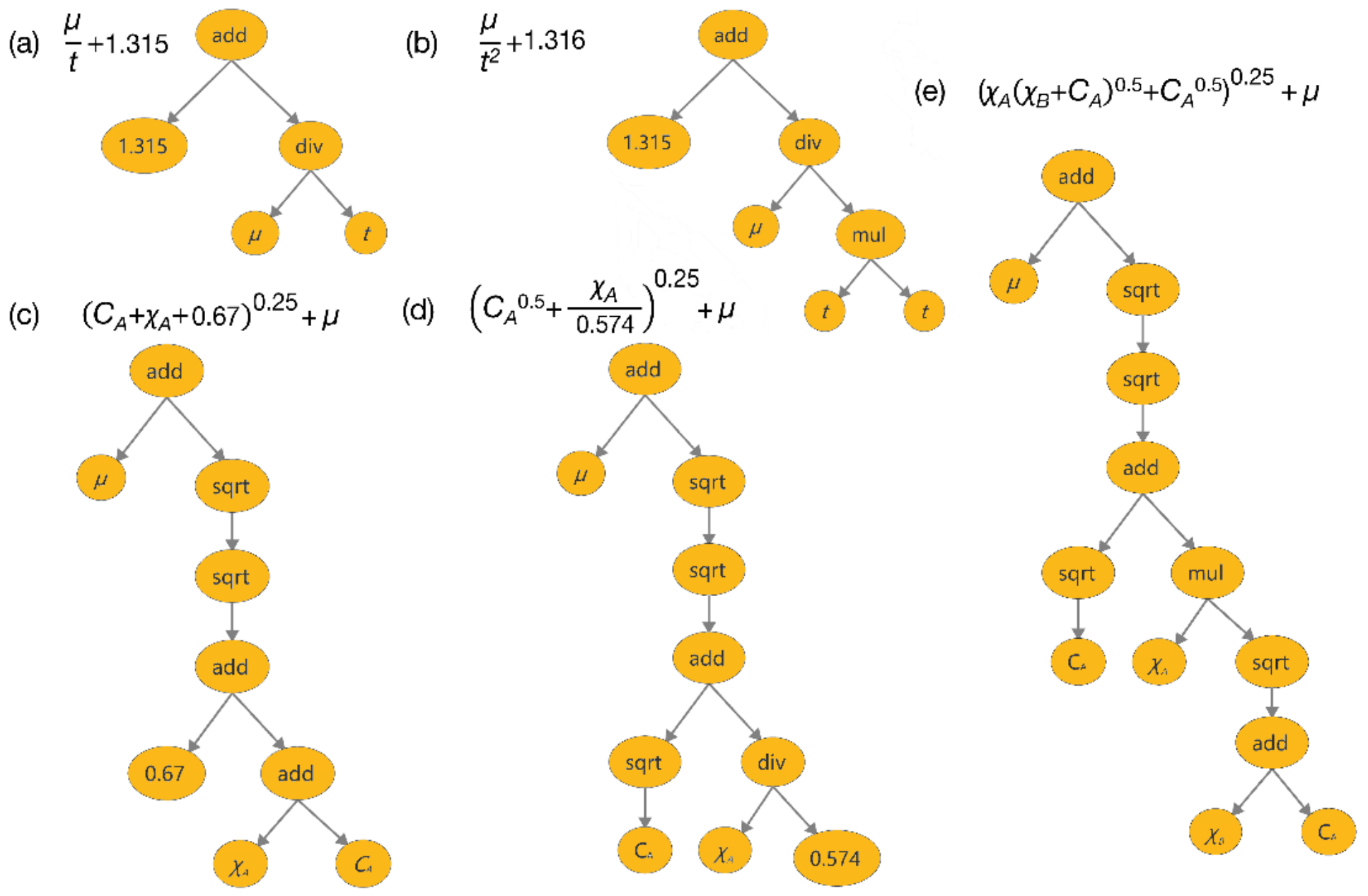

**Figure. S7. The examples of the tree structure for sample analytical formulas. a,** $\mu/t + 1.315$ with complexity 5. **b,** $\mu/t^2 + 1.316$ with complexity 7. **c,** $(C_A+\chi_A+0.67)^{0.25}+\mu$ with complexity 9. **d,** $\left(C_A^{0.5}+\frac{x_A}{0.574}\right)^{0.25}+\mu$ with complexity 10. **e,** $\left((\chi_A(\chi_B+C_A)^{0.5}+C_A^{0.5}\right)^{0.25}+\mu$ with complexity 13.

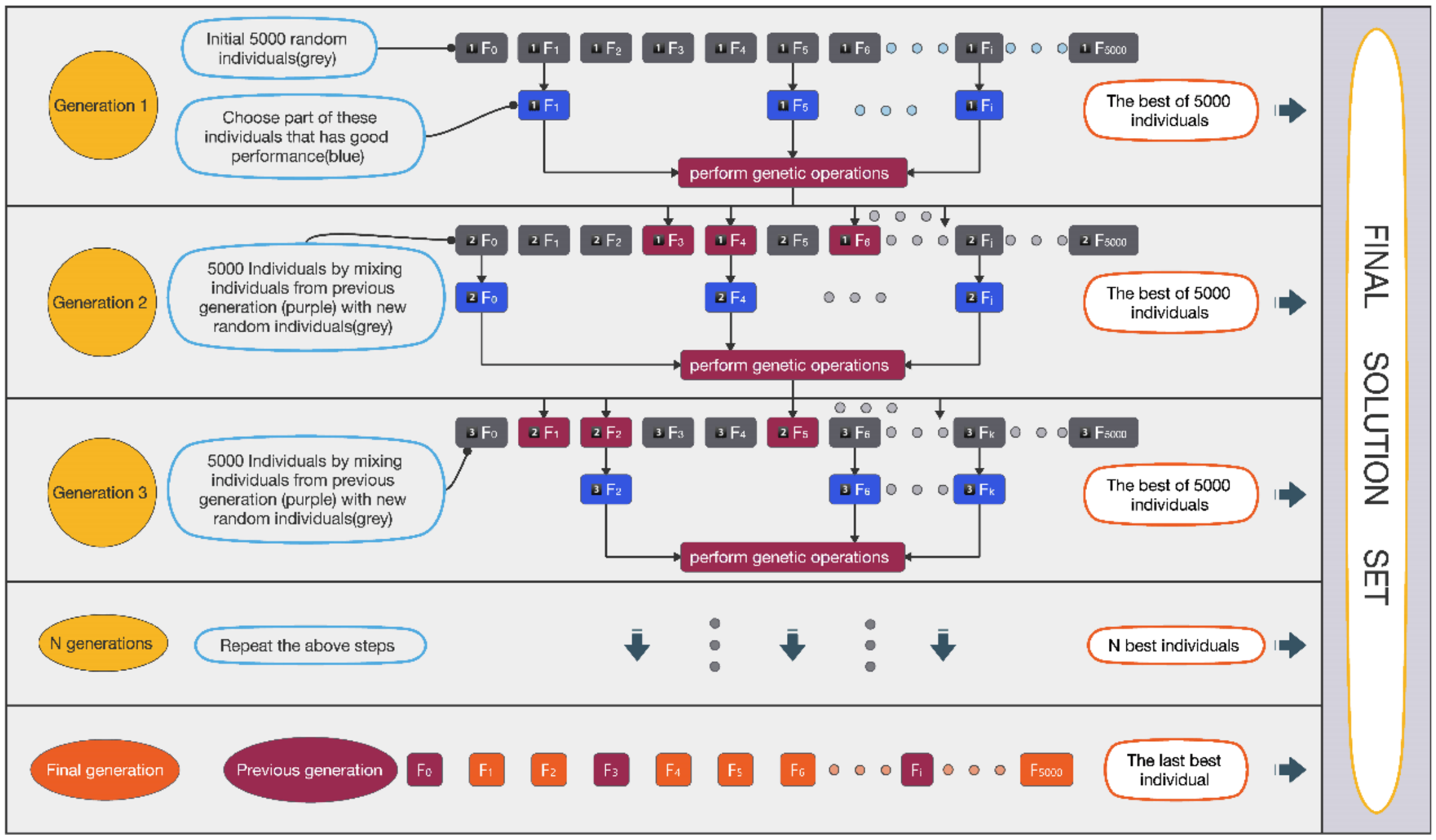

**Fig. S8. Flowchart of symbolic regression based on genetic algorithm**. Each generation includes 4000 randomly generated analytical formulas and 1000 formulas inherited from the previous generation. Each generation provides its best formulas (least MAE) to the final solution sets

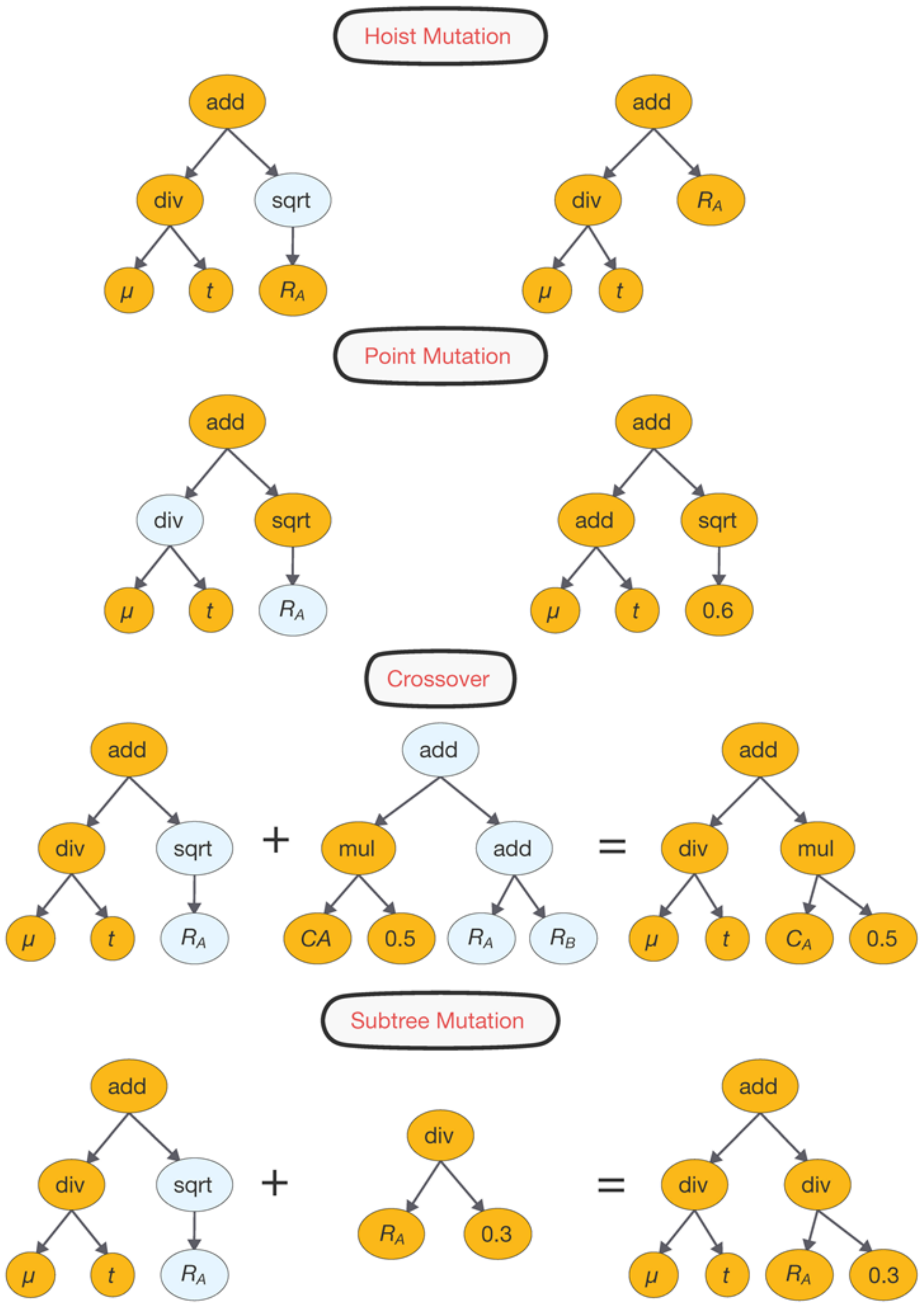

**Fig. S9. The schematic diagrams of four genetic operations used in current genetic programming. a,** The hoist mutation method selects a subtree of a randomly selected subtree from the winner of a tournament and replace the previously selected subtree with it. **b,** The point mutation method randomly selects some nodes from the winner of a tournament and replace it with other building blocks. **c,** The crossover method randomly selects a subtree from the winner of a tournament and replace it with a subtree selected at random from the winner of another tournament. **d,** The subtree mutation method randomly selects a subtree from the winner of a tournament and replaces it with a subtree generated at random.

**Table S1. The nine analytical formulas at the Pareto front in Fig. 2(b).**

| Point | Formulas | MAE (eV) | Number of parameters | Complexity |
|---|---|---|---|---|
| A | $\frac{1.751}{t}$ | 0.0253 | 1 | 3 |
| B | $(\chi_A+1.884)^{0.5}$ | 0.0246 | 1 | 4 |
| C | $(Q_A^{0.5}+1.347)^{0.5}$ | 0.0223 | 1 | 5 |
| D | $(2\chi_A R_A)^{0.5}$ | 0.0219 | 2 | 6 |
| E | $1.612\frac{\mu}{t}+1.073$ | 0.0216 | 2 | 7 |
| F | $(t+0.854+\mu Q_A)^{0.5}$ | 0.0206 | 3 | 8 |
| G | $((tQ_A)^{0.5}+\frac{2.398}{\chi_B})^{0.5}$ | 0.0200 | 3 | 9 |
| H | $\left((\frac{Q_A}{\chi_B})^{0.5}+(2.008+R_B)^{0.5}\right)^{0.5}$ | 0.0181 | 3 | 10 |
| I | $\left((\frac{Q_A}{\chi_B})^{0.5}+(2.76-\frac{t}{R_A}+R_B)^{0.5}\right)^{0.5}$ | 0.0180 | 5 | 14 |

**Table S2. The $V_{RHE}$s of eighteen known oxide perovskite OER catalysts reported in literature at ~ 5 mA cm$^{-2}$ in 0.1 M KOH/NaOH.** For comparison, the available $V_{RHE}$s reported by other groups in literatures are also provided.

| Materials | $V_{RHE}$ (eV) (This work) | $V_{RHE}$ (eV) (Reference) | References |
|---|---|---|---|
| $La_{0.5}Pr_{0.5}FeO_3$ | 1.725 | N/A | *J. Alloys Comp*., 2015, **649**, 1260-1266 |
| $PrFeO_3$ | 1.758 | N/A | *J. Chem. Sci.*, **126**, 517–525 |
| $LaFeO_3$ | 1.758 | 1.78 | *Nano Energy*, 2018, **47**, 199-209 |
| $LaMnO_3$ | 1.788 | 1.80 | *ChemSusChem* 2016, **9**, 1-10 |
| $LaMn_{0.5}Ni_{0.5}O_3$ | 1.722 | N/A | *Phys. Rev. B*, **65**, 184416 |
| $LaNi_{0.8}Fe_{0.2}O_3$ | 1.759 | 1.74 | *J. Mater. Chem. A*, 2015, **3**, 9421-9426 |
| $LaNi_{0.9}Fe_{0.1}O_3$ | 1.790 | 1.77 | *J. Mater. Chem. A*, 2015, **3**, 9421-9426 |
| $Sr_{0.25}La_{0.75}Fe_{0.5}Co_{0.5}O_3$ | 1.718 | 1.76 | *ChemSusChem*, 2015, **8**, 1058-1065 |
| $LaNiO_3$ | 1.707 | 1.66 | *J. Phys. Chem. Lett.,* 2013, **4**, 1254-1259 |
| $LaMn_{0.5}Cu_{0.5}O_3$ | 1.774 | N/A | *AIP Conf Proc*, 2014, **1591**, 1630 |
| $LaCoO_3$ | 1.721 | 1.64 | *Chem. Mater.,* 2014, **26**, 3368-3376 |
| $La_{0.5}Ca_{0.5}CoO_3$ | 1.682 | 1.71 | *Mater Res Bull* 2000, **35**. 1955–1966 |
| $La_{0.8}Sr_{0.2}CoO_3$ | 1.688 | ~1.63 | *Mater. Chem. Phys.*, 1986, **14**, 397-426 |
| $La_{0.4}Sr_{0.6}CoO_3$ | 1.695 | ~1.63 | *Mater. Chem. Phys.*, 1986, **14**, 397-426 |
| $BaFeO_3$ | 1.686 | N/A | *Electrochim Acta*, 2018, **289**, 428-436 |
| $La_{0.2}Sr_{0.8}CoO_3$ | 1.681 | 1.70 | *Int. J. Electrochem. Sci* 2016, **11**, 8633-8645 |
| $SrCoO_3$ | 1.670 | 1.65 | *Nature Chem* 2017 **9,** 457–465 |
| $Ba_{0.5}Sr_{0.5}Co_{0.8}Fe_{0.2}O_3$ | 1.639 | 1.61 | *Science* 2011, **334**, 1383-1385 |

**Table S3. The setup parameters in gplearn for symbolic regression.**

| Parameter | Value |
|---|---|
| **population size** | 5000 |
| **Generations** | 20 |
| **stopping criteria** | 0.01 (eV) |
| **crossover probability(pc)** | 0.5, 0.95 (step = 0.025) |
| **subtree mutation probability (ps)** | (1-pc)/3, (0.92-pc)/3 (step = 0.01) |
| **hoist mutation probability (ph)** | ps |
| **point mutation probability (pp)** | 1-pc-ps-ph |
| **function set** | add, sub, mul, div, sqrt |
| **parsimony coefficient** | 0.0005, 0.0015 (step = 0.0005) |
| **tournament size** | 20 |
| **metric** | mean absolute error (MAE) |
| **constant range** | (-1,1) |

The ‘population size’ means the number of analytical formulas in each generation, the ‘generations’ means the number of generations for each hyper-parameter, the ‘stopping criteria’ is the MAE value that the program stops, the meanings of pc, ps, ph and pp are shown in Fig. S9. The ‘function set’ is the basic building blocks containing mathematical operators, the ‘parsimony coefficient’ is a constant that penalizes large individuals by adjusting their fitness to make them less favorable for selection, the ‘tournament size’ controls the number of individuals in each tournament, the ‘metric’ measures how well an individual fits, the ‘constant range’ is the range of constants included into individuals.

**Data S1. (separate file)**

**Screened oxide perovskites.** The new oxide perovskites are listed in the sequence of μ/t. In the table, as a typical example, 'Cs+0.5La3+0.5Mn3+0.25Mn4+0Fe3+0.75Fe4+0Co3+0Co4+0Ni3+0Ni4+1.0' means '$Cs^{+}_{0.5}La^{3+}_{0.5}Mn^{3+}_{0.25}Mn^{4+}_{0}Fe^{3+}_{0.75}Fe^{4+}_{0}Co^{3+}_{0}Co^{4+}_{0}Ni^{3+}_{0}Ni^{4+}_{1.0}O_3$'

**Data S2. (separate file)**

The ionic and compositional data used for screening of new oxide perovskites. The table shows examples of the construction of twenty-three perovskites oxides.

## References

1. Yiqun Wang, N. W., James M. Rondinelli, Symbolic regression in materials science. *arXiv:1901.04136v2* (2019).
2. Schmidt, M. & Lipson, H, Distilling Free-Form Natural Laws from Experimental Data. *Science* **324**, 81-85 (2009).
3. Koza, J. R. *Genetic Programming: On the Programming of Computers by Means of Natural Selection.* (*MIT Press, Cambridge, MA*). (1992).
4. Forrest, S, Genetic Algorithms - Principles Of Natural-Selection Applied To Computation. *Science* **261**, 872-878 (1993).
5. Butler, K. T., Davies, D. W., Cartwright, H., Isayev, O. & Walsh, A, Machine learning for molecular and materials science. *Nature* **559**, 547-555 (2018).
6. *https://gplearn.readthedocs.io/en/latest/intro.html*.